\documentclass[letter]{aa}   

\usepackage{graphicx}
\usepackage{txfonts}
\usepackage{natbib}
\usepackage{hyperref}
\hypersetup{
    final=true,
    pageanchor=true,
    colorlinks=true,
    breaklinks=true,
    linkcolor=blue,
    citecolor=blue,
    urlcolor=blue,
    pdfpagemode=UseNone}

\usepackage[switch, modulo]{lineno}

\begin{document} 

\title{Near-infrared spectropolarimetry of a $\delta$-spot}

\author{%
    H.\ Balthasar\inst{1},
    C.\ Beck\inst{2},
    R.\ E.\ Louis\inst{1},
    M.\ Verma\inst{1,3}, \and
    C.\ Denker \inst{1}}

\institute{%
    Leibniz-Institut f\"{u}r Astrophysik Potsdam (AIP),
    An der Sternwarte~16,
    14482 Potsdam,
    Germany\\
    \email{hbalthasar@aip.de}
\and
    National Solar Observatory/Sacramento Peak,
    P.O.\ Box~62,
    Sunspot, 
    NM~88349,
    U.S.A.   
\and
    Max-Planck-Institut f{\"u}r Sonnensystemforschung,
    Max-Planck-Stra{\ss}e~2,
    37191 Katlenburg-Lindau,
    Germany}

\date{Received \today; accepted MMMM DD, YYYY}

\abstract{Sunspots harboring umbrae of both magnetic polarities within a common
penumbra ($\delta$-spots) are often but not always related to flares. We present
first near-infrared (NIR) observations (\ion{Fe}{i} $\lambda$1078.3\,nm and
\ion{Si}{i} $\lambda$1078.6\,nm spectra) obtained with the Tenerife Infrared
Polarimeter (TIP) at the Vacuum Tower Telescope (VTT) in Tenerife on 2012
June~17, which afford accurate and sensitive diagnostics to scrutinize the
complex fields along the magnetic neutral line of a $\delta$-spot within active
region NOAA 11504. We examine the vector magnetic field, line-of-sight (LOS)
velocities, and horizontal proper motions of this rather inactive $\delta$-spot.
We find a smooth transition of the magnetic vector field from the main umbra to
that of opposite polarity 
($\delta$-umbra), but a discontinuity of the horizontal
magnetic field at some distance from the $\delta$-umbra on the polarity
inversion line.
The magnetic field decreases faster with
height by a factor of two above the $\delta$-umbra.
The latter is surrounded by its own Evershed flow.
The Evershed flow coming from the main umbra ends at a line dividing the spot 
into two parts. This line is marked by the occurrence of central emission in the 
\ion{Ca}{ii}$\lambda$854.2\,nm line. Along this line, high chromospheric LOS-velocities 
of both signs appear.
We detect a shear flow within the horizontal flux transport velocities parallel to the
dividing line.
}

\keywords{%
    Sun: sunspots --
    Sun: magnetic field --  
    Sun: photosphere --
    Sun: chromosphere --
    Techniques: polarimetric}

\maketitle


\section{Introduction}

In some cases, a sunspot penumbra contains umbrae of opposite magnetic
polarities as first described by \citet{kuenzel1}. A few years later, \citet{kuenzel2}
introduced the new category `$\delta$-spot' extending Hale's spot classification
in terms of the magnetic field.
Although $\delta$-spots are frequent and often
associated with flares \citep[see][]{sammis_etal}, accurate measurements of the
magnetic field vector with high resolution are still rare, especially in the
NIR. In the context of flares and $\delta$-spots, observations beyond 1~$\mu$m
have been limited to white-light flare emission at 1.56\,$\mu$m after major
X-class flares \citep[e.g.,][]{Xu2004, Xu2006} and sporadic long-wavelength infrared 
spectropolarimetry in the \ion{Mg}{i} $\lambda$12.32\,$\mu$m
line \citep[e.g.,][]{Jennings2002}.

Velocity investigations in the visible are more frequent.
Persistent downflows of up to 14\,km\,s$^{-1}$ within the penumbra close to 
the polarity inversion line (PIL) have been reported by \citet{valentin}.
At lower spectral and spatial resolution, similar
downflows of 1.5--1.7\,km\,s$^{-1}$ have been detected by \citet{takizawa} in
regions of penumbral decay.
Evershed flows of opposite polarity umbrae converge near the PIL and have to
pass each other, which is indicative of an interleaved system of magnetic field
lines \citep{lites,hirzi}. Moreover, significant up- and downflows in excess of
10\,km\,s$^{-1}$ are encountered during flares \citep{cathrin}.

The observed $\delta$-spot shares a remarkable visual resemblance with its
counterpart in the flare-prolific active region NOAA 10930, which by contrast
exhibits significant rotation (up to $8^\circ$ hour$^{-1}$) of the
$\delta$-umbra \citep{MinChae}. Shear flows along the PIL potentially contribute
to the build-up of non-potential magnetic field configurations
\citep[cf.,][]{denker_etal}. Conversely, \citet{tan_etal} note that the shear
flow between opposite polarity umbrae 
decreases from 0.6\,km\,s$^{-1}$ before
to 0.3\,km\,s$^{-1}$ after an X3.4 flare.

In this Letter, we introduce NIR spectropolarimetry as a diagnostic tool to
study the intriguing properties of $\delta$-spots.


\section{Observations}

On 2012 June~17, not far from the maximum of the solar cycle No. 24, we observed
a sunspot with a $\delta$-configuration in active region NOAA 11504 at
$18^\circ$\,S and $29^\circ$\,W ($\mu = \cos\vartheta = 0.82$). A map of the
full Stokes vector was obtained with 
the Tenerife Infrared Polarimeter \citep[TIP;][]{tip2} 
and adaptive optics' correction \citep{Berkefeld2010} at the 
Vacuum Tower Telescope \citep[VTT;][]{vonderLuehe1998}.
The dispersion of the \ion{Fe}{i} $\lambda$1078.3\,nm and \ion{Si}{i}
$\lambda$1078.6\,nm spectra amounts to 2.19~pm pixel$^{-1}$.
The map (10:00--10:38\,UT) consists of 180 scan positions with a
step size of 0\farcs36. The image scale of 0\farcs175 pixel$^{-1}$ along the
slit direction was resampled to 0\farcs35 pixel$^{-1}$ to reduce the noise
in preparation for the spectral inversions. For each Stokes parameter, we
accumulated ten exposures of 250\,ms \citep[see][]{hobagoem}.

The \ion{Fe}{i} and \ion{Si}{i} lines sample different atmospheric layers of
the quiet Sun, only in dark umbrae they originate at almost the same height
\citep[see][]{hobagoem}. We derived the magnetic field vector using `Stokes
inversions based on response functions' \citep[SIR;][]{sir}. The two lines were
inverted separately. The magnetic field strength, the magnetic inclination
and azimuth as well as the Doppler velocity were kept constant with height,
whereas we use three nodes for the temperature. Initially, the magnetic azimuth
ambiguity was resolved by selecting directions matching a radial
configuration with a common center \citep[see][]{hoba06}. The subsequently 
applied minimum energy
method of \citet{leka} yields a more reliable configuration.
Finally, we rotated the magnetic vector to
the local reference frame and corrected the images for projection effects
\citep[see][]{verma_etal}. The resulting field-of-view (FOV) of 60~Mm $\times$
40~Mm served as a reference for all auxiliary data.

\begin{figure}[t]                         
\centerline{\includegraphics[width=\columnwidth]{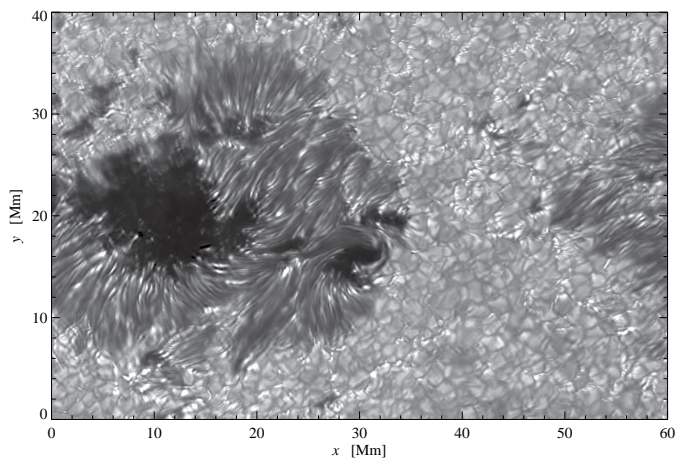}}
\caption{G-band image of the $\delta$-spot in active region NOAA 11504.}
\label{gband_image}
\end{figure}

Multi-frame blind deconvolution \citep[MFBD;][]{vanNoort2005} was applied
to high-resolution (0\farcs04 pixel$^{-1}$) G-band context images (see
Fig.~\ref{gband_image}), which were recorded in the VTT's optical
laboratory. Additional \ion{Ca}{ii} $\lambda$854.2\,nm spectra 
with a dispersion of 0.82~pm pixel$^{-1}$ and an image scale
of 0\farcs35 pixel$^{-1}$ were acquired at the Echelle spectrograph with a
camera mounted next to TIP. Integration times of 9\,s matched the TIP
accumulation cycle. The spectral range also covered the \ion{Si}{i}
$\lambda$853.6\,nm line, which is weak and has a high excitation potential of
6.15\,eV, thus, representing deep photospheric layers. 

\begin{figure}[t]
\centerline{\includegraphics[width=\columnwidth]{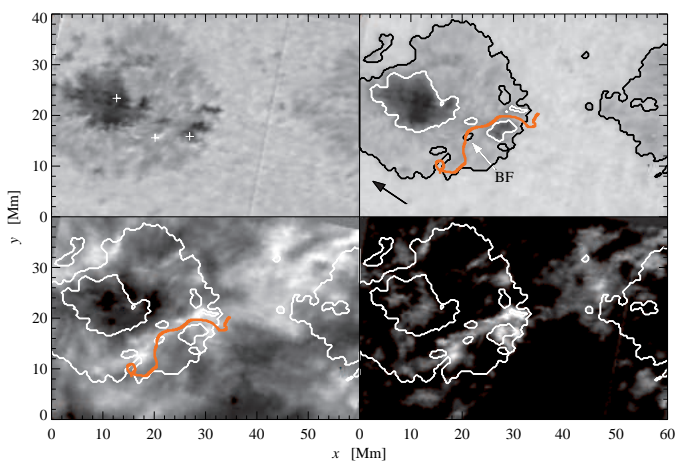}}
\caption{Maps of continuum intensities at $\lambda$1078 and $\lambda$854\,nm,
\ion{Ca}{ii} $\lambda$854.2\,nm line-core intensity, and maximum of the
    central \ion{Ca}{ii} emission (top-left to bottom right). The ordinate is
    along the solar North-South direction. The black arrow points towards
    disk center, and the white arrow indicates the bright feature BF near the PIL
    (thick red line). Contour lines denote the umbral and penumbral boundaries.
    Three white crosses mark the positions of the spectral profiles in
    Fig.~\ref{spectra}.}
\label{cont_image}
\end{figure}

Magnetograms of
the Helioseismic and Magnetic Imager \citep[HMI;][]{hmi} on board of the Solar
Dynamics Observatory \citep[SDO;][]{Pesnell2012} 
accompany the VTT data and provide information of the
magnetic field evolution and 
are used to determine horizontal velocities.


\section{Results}

Penumbral filaments related to the $\delta$-umbra display a counterclockwise 
twist in the central part of Fig.~\ref{gband_image}, while those of the main umbra
show an opposite twist in the vicinity of the $\delta$-umbra. The braided strands of
penumbral filaments are parallel to the PIL and roughly perpendicular to the
line connecting the centers of the main and $\delta$-umbrae.

\begin{figure}[t]
\centerline{\includegraphics[width=\columnwidth]{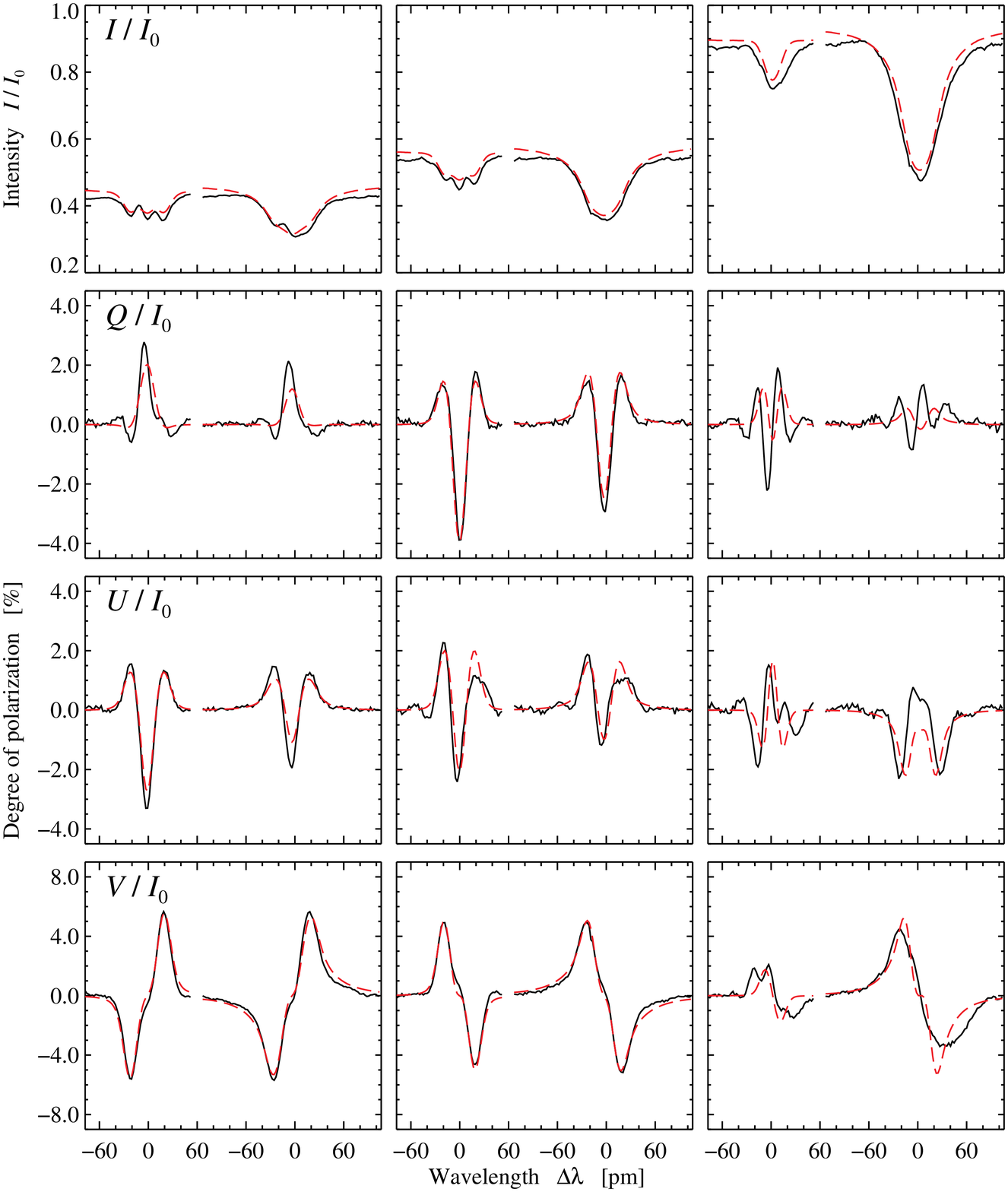}}
\caption{Observed (solid) and inverted (dashed) Stokes-profiles in the umbra of
   the main spot (left), the $\delta$-spot umbra (middle), and in the penumbra
   near BF (right). The wavelength is given with respect to the center of
   the \ion{Fe}{i} $\lambda$1078.3\,nm (left) and \ion{Si}{i}
   $\lambda$1078.6\,nm (right) lines, respectively. $I_\mathrm{0}$ is the
   quiet-Sun continuum intensity at disk center.}
\label{spectra}
\end{figure}

Figure~\ref{cont_image} shows continuum intensity images at $\lambda$1078\,nm
and $\lambda$854\,nm. Apart from the main umbra, several dark features reside in
the common penumbra. Yet only one dark umbral core close to the right border of
the penumbra at the location $(x, y) \sim (28, 15)$~Mm possesses opposite
magnetic polarity. Central emission of the \ion{Ca}{ii} line is mainly present
in the $\delta$-umbra and along a diagonal starting just to the north of the
$\delta$-umbra and extending for about 30~Mm to the south-east. We refer to this
alignment of brightenings as `central emission line' (CEL), which is in some
locations, but not everywhere, co-spatial with the PIL. 
The CEL is inconspicuous in continuum images.

A comparison of observed and fitted Stokes-profiles 
for three selected pixels (marked in Fig.~\ref{cont_image}) 
is shown in Fig.~\ref{spectra}. Observed and fitted profiles agree rather well for the 
main and the $\delta$-umbra, but for a point on the PIL near the bright
feature (BF) indicated in Fig.~\ref{cont_image},
a single-component inversion is not able to reproduce the observed 
multi-lobe profiles.

\begin{figure}[t]
\centerline{\includegraphics[width=\hsize]{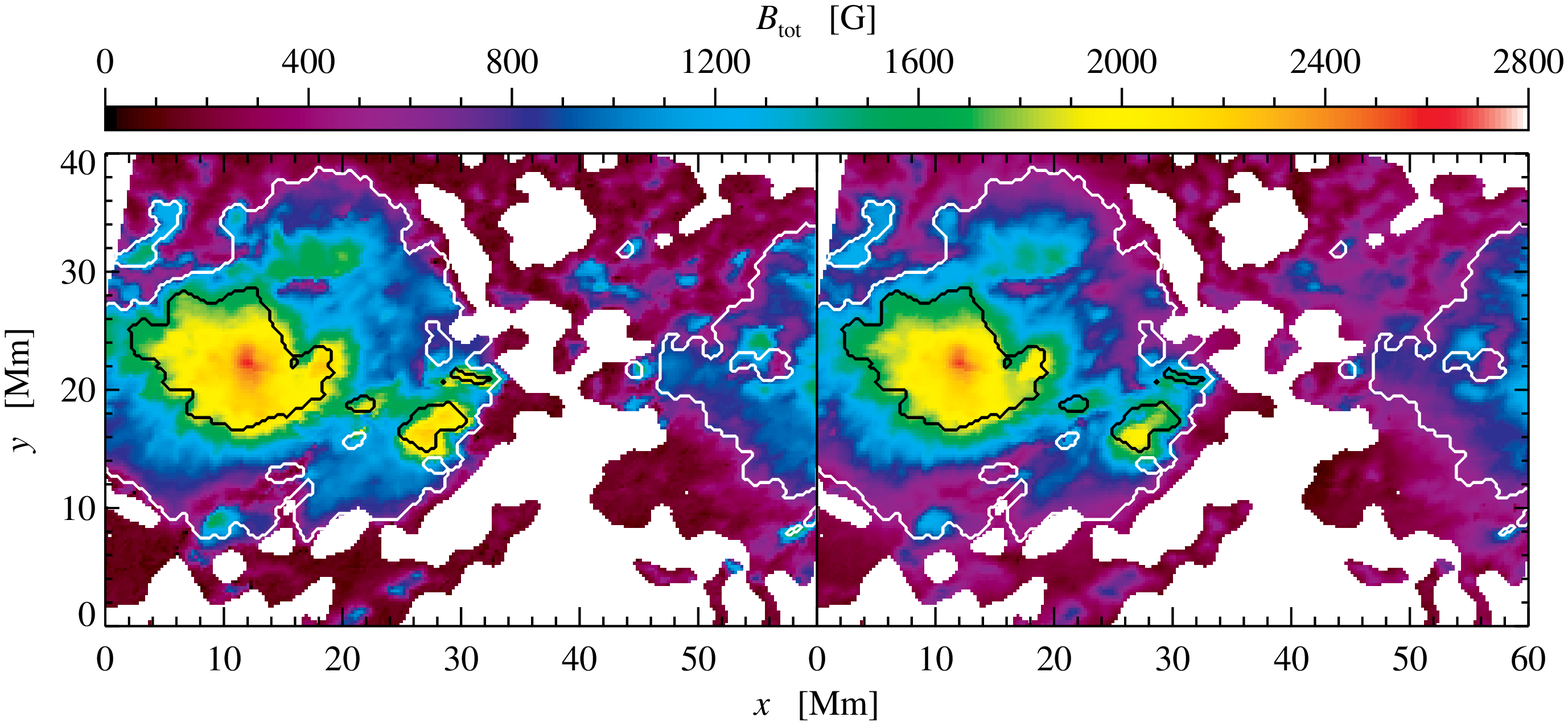}}
\centerline{\includegraphics[width=\hsize]{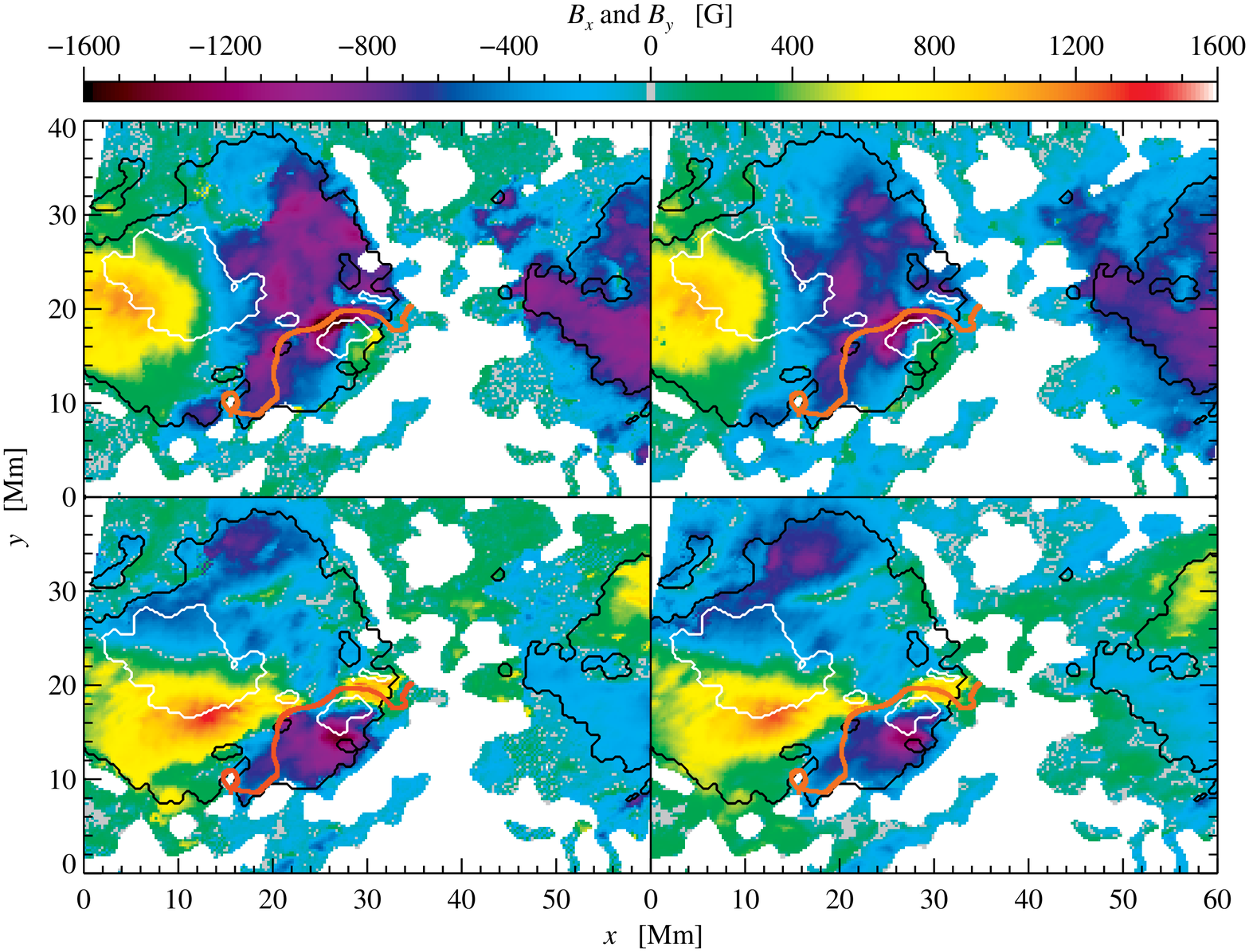}} 
\centerline{\includegraphics[width=\hsize]{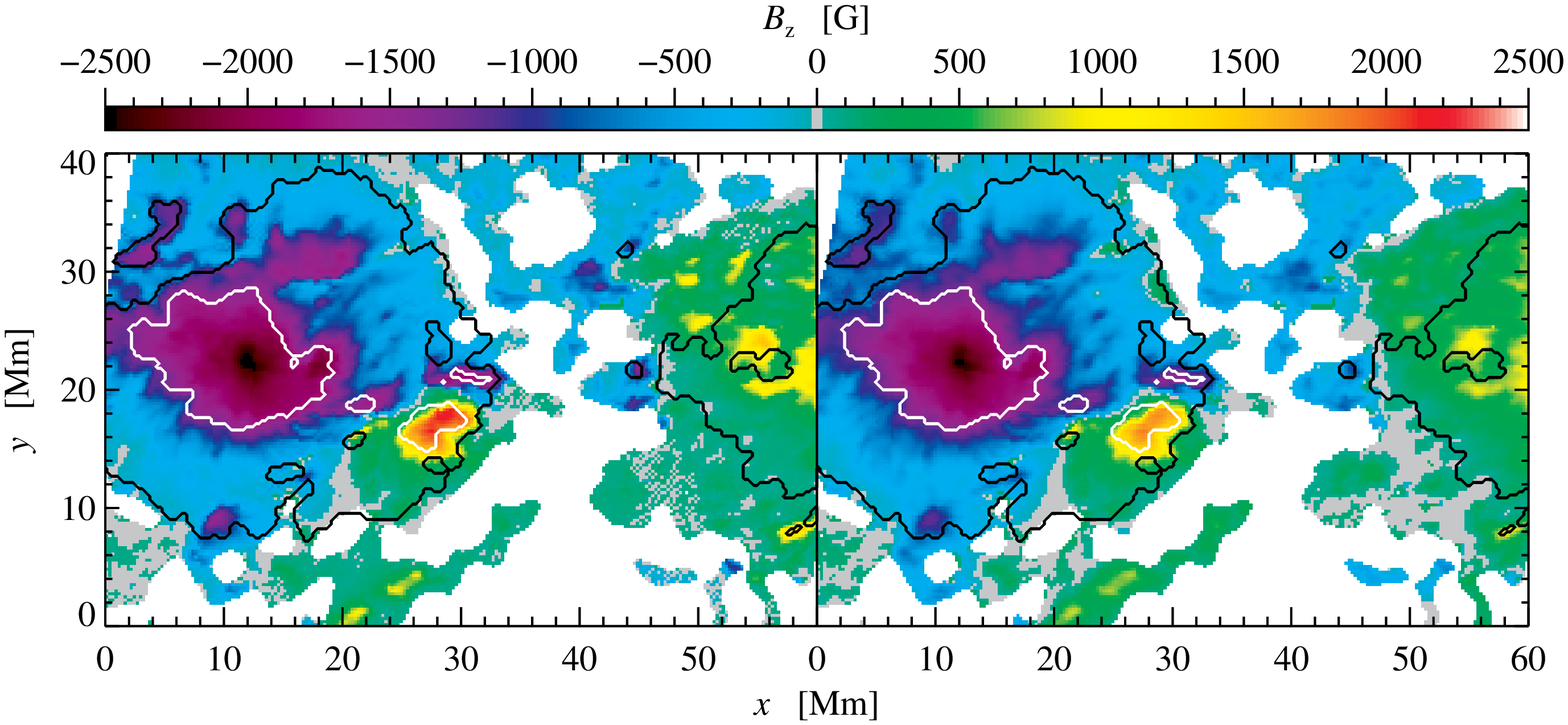}}
\caption{Total magnetic field $B_\mathrm{tot}$ and Cartesian components $B_x$,
    $B_y$, and $B_z$ (top to bottom) derived from inversions of the \ion{Fe}{i}
    $\lambda$1078.3\,nm (left) and \ion{Si}{i} $\lambda$1078.6\,nm (right)
    lines, respectively.}
\label{magn_image}
\end{figure}

The total magnetic field strength $B_\mathrm{tot}= |\vec{B}|$ and Cartesian
components of the magnetic field vector $\vec{B} = (B_x, B_y, B_z)$ have been
determined with SIR for the \ion{Fe}{i} $\lambda$1078.3\,nm  and \ion{Si}{i}
$\lambda$1078.6\,nm lines. The vertical component $B_z$ in Fig.~\ref{magn_image}
reveals clearly the opposite polarities of the main and $\delta$-umbrae. The
parasitic polarity is positive like that of the leading sunspot in the active
region. Inversions of the \ion{Fe}{i} and \ion{Si}{i} lines bring forth field
strengths in the main umbra of up to 2600\,G and 2570\,G, respectively.
Corresponding values in the $\delta$-umbra reach only 2250\,G and 2030\,G. The
height difference between the formation layers of the two NIR spectral lines
is larger above
the $\delta$-umbra, because it is hotter than the main umbra. Nevertheless,
dividing the difference in $B_\mathrm{tot}$ by the difference in height covered 
by the two lines
\citep[see][]{hobagoem} indicates that the magnetic field decreases much faster
with height in the $\delta$-umbra as compared to the main umbra. In the $\delta$-umbra,
the mean formation height of the \ion{Fe}{i} line is 150\,km and that of the \ion{Si}{i}  line is 210\,km. The corresponding values at the center of the main umbra are 140\,km and 180\,km,
respectively. With these values, we find magnetic gradients of 
$\Delta B_\mathrm{tot} / \Delta z \approx $\,$4.5\,\pm\,1.4$\,G\,km$^{-1}$ and 2\,$\pm$\,0.5\,G\,km$^{-1}$, respectively.
$B_\mathrm{tot}$ is rather low below the southeastern part of the CEL.

\begin{figure}[t]
\centerline{\includegraphics[width=\columnwidth]{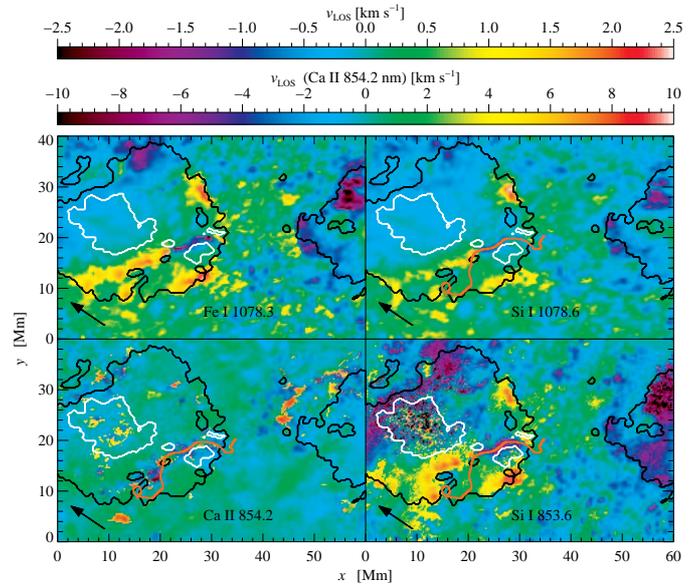}}
\caption{Doppler velocities obtained with SIR (top) based on the the \ion{Fe}{i}
    $\lambda$1078.3\,nm (left) and \ion{Si}{i} $\lambda$1078.6\,nm (right)
    lines, respectively. Line-fitting (bottom) yields LOS velocities of the
    chromospheric \ion{Ca}{ii} $\lambda$854.2\,nm (left) and photospheric
    \ion{Si}{i} $\lambda$853.6\,nm (right) lines. 
    Note the different scale bar for the velocities from the \ion{Ca}{ii} line.
}
\label{velo_image}
\end{figure}

\begin{figure}[t]
\centerline{\includegraphics[width=\columnwidth]{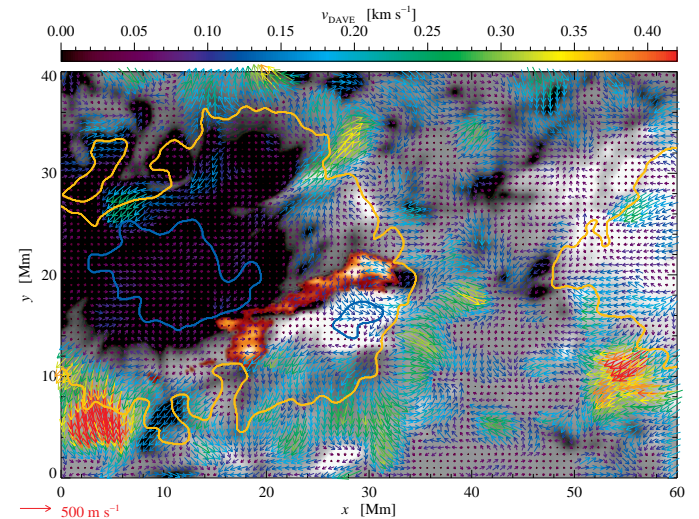}}
\caption{One-hour averaged HMI magnetogram (clipped to $\pm$500\,G in gray scale) with superimposed vectors of $v_\mathrm{DAVE}$.
Length (see arrow below the figure) and color of the vectors represent the magnitude of $v_\mathrm{DAVE}$. Emission along the CEL is indicated by the red background.}
\label{FigDAVE}
\end{figure}

$B_x$ varies smoothly with negative
values in the region between the two umbrae. The perpendicular component $B_y$
has a positive sign south of the main umbra and north of the $\delta$-umbra,
whereas it is negative on the other side. In the northern part of the spot,
these opposite signs are well separated, but in the southern part, at the edge
of BF, large values of $B_y$ with opposite signs occur in close proximity. 
Inside BF, $B_z$ is positive. The PIL bends at BF and separates 
from the CEL.
At these locations, anomalous multi-lobed Stokes-profiles arise (see right column in
Fig.~\ref{spectra}), which cannot be reproduced by a single-component inversion.
(A two-component inversion is beyond the scope of this Letter.)
Changing the magnetic azimuth by $180^\circ$ even compounds this problem.
\citet{lites} explained such profiles in terms of interleaved systems of
magnetic field lines. Such interleaved systems might facilitate reconnections,
but in our case, the region was small enough not to cause a major flare.

The Doppler velocities derived from different lines are summarized in
Fig.~\ref{velo_image}. Photospheric lines exhibit the Evershed effect as
blue-shifts in the upper left part of the spots and as red-shifts on the right
side. The Evershed effect is most pronounced in the \ion{Si}{i}
$\lambda$853.6\,nm line that forms at deep layers. 
The Evershed flow from the main umbra is interrupted in proximity 
to the area where it reaches
the CEL, and where the filaments coming from the main umbra end.
Beyond this line, it drops and finally reaches another maximum at
the outer edge of the penumbra.
A strong blue-shift of about 1\,km\,s$^{-1}$ is
apparent just to the north of the $\delta$-umbra along the PIL. We cannot 
finally decide
whether this blue-shift corresponds to an upflow or an Evershed-like flow away
from the $\delta$-umbra because the spot is at some distance from disk center.
This blue-shift encompasses the PIL, where the magnetic field is
horizontal, implying that the flow is more horizontal as well. 
The Evershed effect is also present in the \ion{Ca}{ii} $\lambda$854.2\,nm line.
Patches of high velocities of up to 8\,km\,s$^{-1}$ exist
along the CEL occurring both as blue-shifts and as red-shifts. 
High red-shifts appear also in the periphery of the spot \citep[cf.,][]{hvar}.
Because these patches are small, we interpret them 
as vertical flows, in agreement with downflows found by \citet{valentin},
although we do not find supersonic flows in photospheric layers
\citep[see][]{takizawa}.

The `differential affine velocity estimator' \citep[DAVE,][]{Schuck2005,
Schuck2006} estimates the magnetic flux transport velocity $v_\mathrm{DAVE}$ 
by utilizing an
affine velocity profile and minimizing the deviation in the magnetic induction
equation. The input HMI magnetograms have been aligned, corrected for geometrical
foreshortening, and adjusted for the $\mu$-dependence of the magnetic field
strength assuming that the field lines are perpendicular to the solar surface.
DAVE depends on the temporal derivative of the magnetograms implemented as a
five-point stencil, spatial derivatives carried out with the \citet{Scharr2007}
operator, and a sampling window of prescribed size (11~pixels $\sim$ 3520\,km).
Arrows overlaid on the magnetogram in Fig.~\ref{FigDAVE} indicate the one-hour
averaged (09:30--10:30~UT) $v_\mathrm{DAVE}$
in the $\delta$-umbra's neighborhood. 
The horizontal velocity structure differs from a regular
sunspot. The most distinct velocity feature is a counterclockwise spiral motion
centered between the $\delta$-umbra and a neighboring dark feature.
In addition, shear flows parallel to the CEL
prevail locally with an imbalance of 0.2\,km\,s$^{-1}$ towards stronger flows on the 
$\delta$-umbra side. This value compares to the 0.3\,km\,s$^{-1}$ found by 
\citet{tan_etal} after a flare. \citet{denker_etal} discussed that such velocities 
do not necessarily built up magnetic shear, and the observed drift does not 
destabilize the magnetic configuration.
Between main and $\delta$-umbra, $v_\mathrm{DAVE}$
amounts to 0.2\,km\,s$^{-1}$
and cross the PIL in the opposite direction as $v_\mathrm{LOS}$.

The CEL seems to be of much higher importance for this spot 
than the PIL. An explanation could be that new bipolar flux emerged southwest of 
the main umbra, and the CEL separates the new flux from the pre-existing one.
Between main and $\delta$-umbra, PIL and CEL are identical, but then the PIL
bends at BF and 
crosses the new bipolar flux. Such a scenario also explains that the Evershed flow 
related to the main umbra ends at the CEL. In the range of repelling field lines
from the two flux systems,
chromospheric brightenings occur as well as between attracting field lines.


\section{Conclusions}

The magnetic field strength $B_\mathrm{tot}$ at deep 
photospheric layers amounts to 2250\,G in the $\delta$-umbra, somewhat less 
than the 2600\,G of the main umbra. It decreases twice as fast with height in the 
$\delta$-umbra compared to the main umbra.
We find a smooth transition of the magnetic vector field between the two umbrae
of opposite polarity. At some distance from the $\delta$-umbra, near BF, we find 
a discontinuity of the horizontal magnetic field due to an interleaved 
system of magnetic field lines.

The $\delta$-umbra causes its own Evershed flow which 
is separated from that of the main umbra.
Along the CEL, we find flows of $\pm$8\,km\,s$^{-1}$ in chromospheric layers.
We observe a
horizontal flow of magnetic features (`magnetic flux transport velocities')
on one side parallel to the CEL with an imbalance of 0.2\,km\,s$^{-1}$.
The magnetic configuration of the $\delta$-spot is seemingly stable for at 
least 10 hours. The group produced an M1.9-flare on June 14 and several C-flares 
during the previous day, but no flare of class C or larger was recorded 
by the GOES satellite during 
19 hours before and 6 hours after our observation. 

The central emission patches in the \ion{Ca}{ii} line at 854.2\,nm 
indicate that important physical processes happen in the
chromospheric layers above $\delta$-spots. For future observations, e.g. with
the GREGOR Infrared Spectrograph \citep[GRIS,][]{Collados2012} at the GREGOR 
solar telescope \citep{GREGOR}, the chromospheric magnetic field should be
investigated too, e.g., to search for probable current sheets.


\begin{acknowledgements} We are grateful to L. Bellot Rubio for carefully reading 
the manuscript and his valuable comments.
The VTT is operated by the Kiepenheuer-Institut f{\"u}r
Sonnenphysik (Germany) at the Spanish Observatorio del Teide of the Instituto de
Astrof{\'\i}sica de Canarias. SDO/HMI data are provided by the Joint Science
Operations Center -- Science Data Processing. MV expresses her gratitude for the
generous financial support by the German Academic Exchange Service (DAAD) in the
form of a PhD scholarship. CD and REL were  supported by grant DE 787/3-1 of the
German Science Foundation (DFG).
\end{acknowledgements}


\end{document}